\documentclass[sigplan,screen]{acmart}
\usepackage{multirow}
\usepackage{multicol}
\usepackage{color}

\AtBeginDocument{%
  \providecommand\BibTeX{{%
    \normalfont B\kern-0.5em{\scshape i\kern-0.25em b}\kern-0.8em\TeX}}}

\setcopyright{acmcopyright}
\copyrightyear{2020}
\acmYear{2020}
\acmDOI{10.1145/1122435.1122456}

\begin{document}

\title{Towards the Datasets Used in Requirements Engineering of Mobile Apps}
\subtitle{Preliminary Findings from a Systematic Mapping Study}

\author{Chong Wang, Haoning Wu}

\affiliation{%
   \institution{School of Computer Science, Wuhan University}
  \city{Wuhan}
   \country{China}
 }
 \email{cwang@whu.edu.cn}

  \author{Peng Liang}
 \affiliation{%
   \institution{School of Computer Science, Wuhan University}
   \city{Wuhan}
   \country{China}
   }
\email{liangp@whu.edu.cn}

 \author{Maya Daneva}
 \affiliation{%
   \institution{EEMCS, University of Twente}
   \city{Enschede}
   \country{The Netherlands}
   }
\email{m.daneva@utwente.nl}

 \author{Marten van Sinderen}
 \affiliation{%
   \institution{EEMCS, University of Twente}
   \city{Enschede}
   \country{The Netherlands}
   }
\email{m.j.vansinderen@utwente.nl}




\renewcommand{\shortauthors}{xxxx, et al.}

\begin{abstract}
\textbf{[Background]} Research on requirements engineering (RE) for mobile apps employs datasets formed by app users, developers or vendors. However, little is known about the sources of these datasets in terms of platforms and the RE activities that were researched with the help of the respective datasets. 
\textbf{[Aims]} The goal of this paper is to investigate the state-of-the-art of the datasets of mobile apps used in existing RE research. 
\textbf{[Method]} We carried out a systematic mapping study by following the guidelines of Kitchenham et al.
\textbf{[Results]} Based on 43 selected papers, we found that Google Play and Apple App Store provide the datasets for more than 90\% of published research in RE for mobile apps. We also found that the most investigated RE activities - based on datasets, are requirements elicitation and requirements analysis.  
\textbf{[Conclusions]} Our most important conclusions are: (1)  there is a growth in the use of datasets for RE research of mobile apps since 2012, (2) the RE knowledge for mobile apps might be skewed due to the overuse of Google Play and Apple App Store, (3) there are attempts to supplement reviews of apps from repositories with other data sources, (4) there is a need to expand the alternative sources and experiments with complimentary use of multiple sources, if the community wants more generalizable results. Plus, it is expected to expand the research on other RE activities, beyond elicitation and analysis.
\end{abstract}




\maketitle

\section{Introduction}

The rapid growth of mobile applications (apps for short) and the mobile app market have resulted in massive sets of app data provided by users, developers, and vendors. For example, users posted their feedback in app repositories, forums, or social networks (e.g. Twitter), while app vendors or developers provide textual descriptions for each app and deliver new releases with the corresponding release notes in app repositories. In recent years, many researchers have recognized these types of app data as an important source to explore requirements engineering (RE) in mobile apps~\cite{Ref2}~\cite{Ref7}. However, a structured account on the sources and types of app data forming the datasets employed in those studies for RE purpose is still lacking. To address the gap, this paper conducts a systematic mapping study (SMS) to report the state-of-the-art the datasets used in RE of mobile apps. By implementing the guidelines proposed by Kitchenham et al.~\cite{Ref3}, we analyzed empirical evidence published in literature using datasets of mobile apps in order to consolidate the understanding of this topic and map out possible directions for future research. 

As a preview of our research on the datasets used in RE of mobile apps, this paper consolidates the empirical publications that talk about datasets consisting of various types of app data for RE purposes. Our paper does this from two perspectives. First, it summarized the sources and types of app data that form the datasets used in the RE literature. Second, it indicated the RE activities in which much research work was focused while using the datasets of mobile apps.  

The rest of this paper is structured as follows: Sect. 2 provides the related work. Sect. 3 presents the research questions and the research process specified in this SMS. Sect. 4 and 5 are the preliminary results of this SMS and our discussion on these results, respectively. Sect. 6 discusses the limitations, followed by the conclusions and future work in Sect. 7. 

\section{Related Work}
There are four literature reviews~\cite{Ref2}~\cite{Ref6}~\cite{Ref4}~\cite{Ref5} relevant to this SMS. Wang et al.~\cite{Ref2} focused on the user feedback provided by the crowds and employed for RE purpose. These authors reported six pieces of metadata that characterized user feedback and that were employed in five RE activities defined in~\cite{Ref1}. Considering the datasets used in the primary studies in the work of Wang et al.~\cite{Ref2}, the sources of data were either mobile apps or other types of software, such as open source software in SourceForge.net. A recent work~\cite{Ref6} examines the usage of data papers published in the MSR (Mining Software Repositories) proceedings in terms of use frequency, users and user purpose. Different to our work, the scope of the data sources in ~\cite{Ref6} were extended to various types of software and limited to publications in one specific conference (MSR). Santos et al.~\cite{Ref4} proposed an overview of the classification techniques or methods in the literature on crowd-based RE, in order to understand the current methods used in the classification of user feedback. Similarly, the authors neither narrowed the user feedback as the app data nor concentrated on their support in which RE activity. Moreover, Laura et al.~\cite{Ref5} introduced the so-called Hall of Apps, a dataset containing top chart's app metadata and reviews collected from Google Play from 4 countries within 30 weeks. Those apps are the top 100 apps available in the top charts of 33 Google Play categories and the Editor Choice list. However, they did not explore the use of this dataset in RE of mobile apps. 
  
\section{Research Process}
\label{process}

\subsection{Research Questions}
\label{RQ}
The objective of this SMS is to investigate the state-of-the-art of the datasets of mobile apps used in existing research, from the perspective of RE. To this end, we formulated two research questions (RQs):

\textbf{RQ1: }\textit{Which data sources are investigated to form the datasets of mobile apps, according to published empirical studies?} 

\textbf{Rationale:} Data referring to different aspects of mobile apps have been collected from various platforms and used in RE of mobile apps. We want to know which data were deemed relevant to mobile apps and were collected from which sources for the purpose of using as the datasets for RE of mobile apps. The answer of this RQ would help us understand the pieces of information and its sources which the researchers and practitioners are working with for the RE purpose.

\textbf{RQ2:} \textit{What are the purposes of using the datasets of mobile apps in RE?} 

\textbf{Rationale:} The datasets of mobile apps can be used for different purposes. We want to know those RE activities that employ the datasets to achieve a specific goal in RE of mobile apps. Answering this RQ would help us understand how useful the datasets of mobile apps are in regard to the five essential RE activities defined in SWEBOK \cite{Ref1}.

\subsection{Study Search}
\label{study_search}
We employed an automatic search in the IEEE digital library to identify studies on this topic. According to our research topic and the RQs, the following search query was created by joining keywords with possible synonyms. 

\textit{TITLE (app OR 'mobile application') AND ABSTRACT (requirement OR feature OR RE)}

Meanwhile, we scoped the time period of the related publication from January 2008 to December 2019, for two reasons: first, both the Apple App Store and Google Play were launched in 2008; second, our automatic search was conducted in March 2020, and the search output published in 2020 is incomplete for this SMS. As a result, we got 639 papers from IEEE.

\subsection{Study Selection}
In this subsection, we defined selection criteria to make the study selection results as objective as possible. Our inclusion (IC) and exclusion criteria (EC) are listed in Table~\ref{tab_ICEC}. 

To evaluate the selection criteria, the first and the second authors conducted a pilot selection using 50 papers, which were randomly selected from those 639 paper returned from IEEE. These two authors reviewed the titles, abstracts, and full texts of those 50 papers independently, and finally included 7 and 8 papers respectively (only one paper is different). After discussion, they got 100\% agreement on including 7 papers as the primary studies of the pilot selection. 

\begin{table} [t]
	\caption{Inclusion and exclusion criteria.}
	\label{tab_ICEC}
	\centering
	\begin{tabular}{cp{7cm}}
		\toprule
		\textbf{No.} & \textbf{Description}\\
		\midrule
		IC1 & The paper directly addresses the topic of RE of mobile apps. \\
		IC2 & The paper addresses both RQs. \\
		IC3 & The paper addresses the datasets used for RE of mobile apps, including functional requirements and quality requirements as specified in ISO 25010. \\
		IC4 & The paper is published in a peer-reviewed journal, conference, or workshop. \\		
		\midrule
		EC1 & The paper addresses the design and/or implementation of a specific mobile application.\\
		EC2 & The paper is not focused on RE of mobile apps.\\
		EC3 & The dataset in the paper is not directly used in RE of mobile apps.\\
		EC4 & The paper is only a research plan or literature review. \\
		EC5 & The full paper is not written in English.\\
		EC6 & The full paper is not available for download.\\
		\bottomrule
	\end{tabular}
\end{table}

\begin{figure}[h]
	\includegraphics[width=\linewidth]{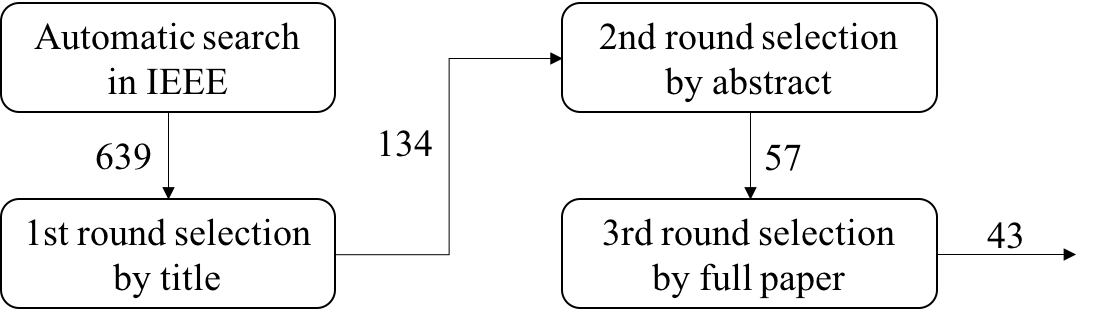}
	\caption{Study search and results in this SMS.}
	\label{selection}
\end{figure}

As Figure~\ref{selection} shows, 639 publications returned from IEEE were manually reviewed for the 1st round selection. This resulted in excluding 505 papers and subjecting 134 papers to the 2nd round selection. The second round selection further excluded 67 papers by reviewing their abstracts and left 57 papers as the input of the 3rd round selection. After carefully reading the full texts of these 57 papers, the 3rd round resulted in the final set of 43 papers as primary studies of this SMS, which is available at https://tinyurl.com/yysyzf5c.

The aforementioned three-round selection process was conducted by the first two authors. The second author performed the 1st and 2nd rounds selection independently, by leaving the uncertain papers to the next round. During the 3rd round selection, when the second author was not confident in the exclusion/inclusion of certain papers, he discussed with the first author to get a consensus on excluding/including them. As soon as the 3rd round selection was done by the second author, the first author randomly selected 20\% of these included papers and gave them careful reading for cross-validating the included papers, and got 100\% agreement on these selected papers.

\subsection{Data Extraction}
\label{extraction}
To answer our RQs (Section~\ref{RQ}), we specified the following extraction rules (EL) to extract data items from the 43 primary studies.
\begin{itemize}
	\item {\textbf{EL1:} To get the overview of primary studies, the publication type and year need to be extracted from the publication information of each primary study.}
	\item {\textbf{EL2:} To answer RQ1, which types of app data and where to collect those data of mobile apps need to be extracted from the full text of this study.}
	\item {\textbf{EL3:} To answer RQ2, which RE activity/activities was/ were mentioned in this study need to be extracted from the full text of this study.}
\end{itemize}

\section{Results}
We performed this SMS according to the steps described in Section~\ref{extraction}. In this section, we report the results to answer each of our RQs defined in Section~\ref{RQ}.

\subsection{Demographics of Primary Studies} 
This section presents the distribution of our 43 included papers per publication types and years.
We found that 72.1\% (31 out of the included 43 studies) are conference papers. Workshop papers account for 18.6\% of the included 43 studies. Whereas, journal papers form 9.3\% (four studies).

\begin{figure}[h]
	\includegraphics[width=\linewidth]{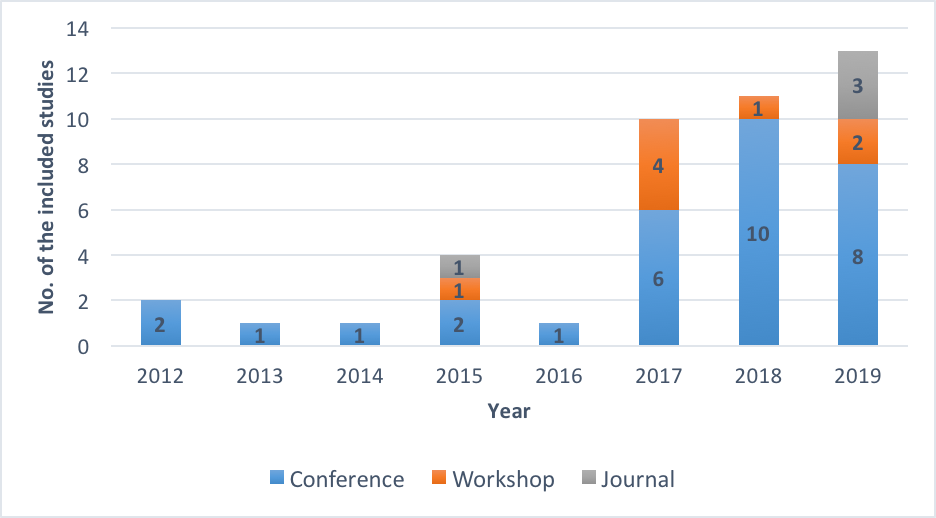}
	\caption{Distribution of the included studies over years.}
	\label{fig_year}
\end{figure}

Furthermore, Figure~\ref{fig_year} shows the number of publications per year. We found that there was no research output published until 2012, and the number of studies using data for RE purpose increased rapidly in recent three years.

\subsection{Data Sources Forming the Dataset of Mobile Apps (RQ1)}
This subsection reports the distribution of data and data sources forming the datasets of mobile apps for RE purpose. 

\begin{table*}[t]
	\caption{Distribution of the selected studies over data sources.}
	\label{tab_datasource1}
	\centering
	\begin{tabular}{p{3.5cm}p{3.5cm}p{1.2cm}p{8cm}}
		\toprule
		\textbf{Category of data source (no. of studies)} & \textbf{Data source} & \textbf{No. of studies} &\textbf{Studies}\\
		\midrule
		\multicolumn{1}{l}{\multirow{5}{2cm}{App repository (38)}} & Google Play & 29 & S2, S3, S4, S6, S7, S9, S11, S16, S18, S19, S20, S21, S23, S24, S26, S27, S28, S30, S31, S32, S33, S34, S36, S37, S38, S39, S40, S42, S43 \\ 
		\multicolumn{1}{l}{~} & Apple App Store & 18 & S4, S8, S9, S11, S15, S17, S18, S19, S25, S29, S30, S33, S34, S35, S39, S41, S42, S43 \\
		\multicolumn{1}{l}{~} & BlackBerry App World & 2 & S5, S13 \\
		\multicolumn{1}{l}{~} & F-Droid & 2 & S2, S3 \\
		\multicolumn{1}{l}{~} & Samsung Android App & 1 & S13 \\ 
		\midrule
		\multicolumn{1}{l}{\multirow{2}{2cm}{Website (2)}} & Stack Overflow & 1 & S10 \\
		\multicolumn{1}{l}{~} & Amazon & 1 & S42 \\
		\midrule   
		Social media (3) & Twitter & 3 & S6, S12, S40 \\
		\midrule
		Questionnaire (2) & Questionnaire & 2 & S14, S22 \\
		\midrule
		App vendor (1) & MyTracks & 1 & S1 \\ 
		\midrule
		~ & \textit{Total} & 60 \\ 
		\bottomrule
	\end{tabular}
\end{table*}

First, we explored the data sources which researchers and practitioners used to collect various types of app data. As shown in Table~\ref{tab_datasource1}, six categories of data sources were reported to provide data of mobile apps to support RE activities. More specifically, 88.37\% of the included studies (38 out of 43 studies) collected apps' data from app repositories to form the datasets used in RE of mobile apps. When zooming in the concrete app repositories in the 2nd column of Table~\ref{tab_datasource1}, it was observed that Google Play and Apple App Store are the two most popular app repositories in our primary studies: 29 out of 38 studies included app data from Google Play, and 18 used Apple App Store. We note that some studies used both repositories. Besides, other app repositories, such Black Berry App World (two studies), F-Droid (two studies), and Samsung Andriod App (one study), were explored by fewer researchers and/or practitioners. Plus, it was observed that some studies had investigated other data sources to get app data for RE purpose. For example, two studies (S10 and S42) collected app data from websites, three studies (S6, S12, and S40) employed app data from social networking website Twitter, two studies (S14 and S22) got app data from questionnaires, while one study (S1) collected app data from the app vendor. 

Furthermore, Table~\ref{tab_datasource2} shows the types of app data identified from the included 43 studies. We found that nine types of app data were explored to support RE activities for mobile apps in those 43 studies. More specifically, app reviews were the most frequently used app data, and the text content of app reviews were analyzed in 33 out of the 43 studies for RE purpose. Four studies analyzed the source code of apps, three studies complemented (S6 and S40) or transfered (S12) the text content analysis with/to the source codes of mobile apps, two studies employed app descriptions, and two studies explored questionnaires on apps. 

\begin{table} [t]
	\caption{Types of app data identified in the selected studies.}
	\label{tab_datasource2}
	\begin{tabular}{p{2.5cm}p{1.2cm}p{3.7cm}}
		\toprule
		\textbf{Type of data} & \textbf{No. of Studies} & \textbf{Studies}\\
		\midrule
		App review & 33 & S1, S2, S4, S6, S7, S8, S9, S11, S16, S17, S18, S19, S20, S23, S24, S25, S27, S28, S29, S30, S31, S32, S33, S34, S35, S36, S37, S38, S39, S40, S41, S42, S43 \\
		App source code & 4 & S5, S13, S21, S26 \\
		Twitter & 3 & S6, S12, S40 \\
		App description & 2 & S21, S34 \\
		Questionnaire & 2 & S14, S22 \\
		Stack Overflow Post & 1 & S10 \\
		App changelog & 1 & S15 \\
		Previous version & 1 & S36 \\
		Commit message & 1 & S3 \\
		\midrule
		\textit{Total:} & 48 & ~ \\
		\bottomrule
	\end{tabular}
\end{table}

In addition, very few studies employed the other five types of app data. For example, S10 reported that app relevant posts in Stack Overflow can be used for RE in mobile apps. Whereas, S15 employed official app changelogs, S36 complemented the analysis on app reviews with the source codes of the previous app releases, and S3 explored commit messages of apps from Github for RE purposes. 

Plus, Table~\ref{tab_datasource2} indicates that 38 out of the included 43 studies employed only one type of app data in RE activities. Whereas, five studies (i.e., S6, S21, S34, S36, and S40) investigated two types of app data in their research.

\subsection{Types of RE Activities Using the Datasets of Mobile Apps (RQ2)}
\label{RQ2}

In this subsection, we applied the data extraction rules in Section~\ref{extraction} to identify the RE activities mentioned in the included 43 studies. The five RE activities are defined in \cite{Ref1}. 

\begin{figure*}[t]
	\includegraphics[width=0.88\linewidth]{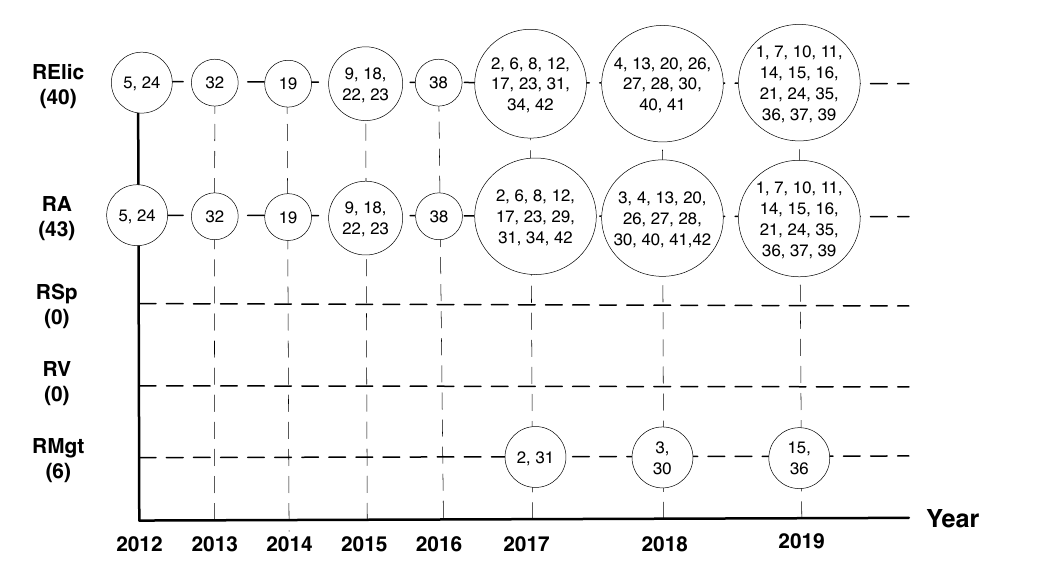}
	\caption{Distribution of the included studies over RE activities and years.}
	\label{fig_RE}
\end{figure*}

Figure~\ref{fig_RE} shows the distribution of the included 43 studies over RE activities and time period, where the number in the bracket under the name of each RE activity denotes the number of included studies that explicitly reported this type of RE activity. We found that requirements analysis (RA) was the most reported RE activity, covering all the 43 included studies. For example, S1 and S43 reported `requirements prioritization' and `classification of NFRs' as one of their keywords, and these two sub-activities are included in RA. Meanwhile, RA can be derived from the aims of S14 -- \textit{`Aims of the conducted survey study were to determine, prioritise and analyze requirements for applications in the field of prevention and health promotion on the basis of the KANO model of customer satisfaction.'}

Regarding requirements elicitation (RElic), it was the second most reported RE activity, accounting for 40 out of the 43 studies. In S12, for example, \textit{`informative tweets are the ones that either express a requirement or offer support for fulfilling a requirement'}, so that topic modeling is used to \textit{`extract the in-common topics in a set of tweets'} as requirements. Whereas, RElic was reported as one of the keywords in S13. 

In addition, we found that requirements management (RMgt) was identified in only six studies. For example, S31 aimed to \textit{`extract from user reviews feedback relevant from a maintenance perspective and suggest the location of such changes in the source code.'} Plus, there is no primary study reporting either requirements specification (RSp) or requirements validation (RV).

Furthermore, we observed that only two studies (S29 and S43) reported only one RE activities -- RA. Around 95.3\% of the included studies covered at least two RE activities. Particularly, 35 out of the 43 studies used app data to support both RElic and RA, five studies (i.e., S2, S15, S30, S31 and S36) reported three RE activities -- RElic, RA, and RMgt, and S3 is the only one study that mentioned both RA and RMgt.  

\section{Discussion}

\subsection{Overview of Primary Studies}
The preliminary results of this survey indicates that the RE of mobile apps is a rising research topic. We first concluded this from the growth of publications (see Figure~\ref{fig_year}) since 2012, especially from 2017 to 2019. Then, this conclusion could be derived from the distribution of primary studies over publication types. It was observed that around 90\% of the included studies were published in conferences or workshops. The reason could be that researchers preferred to share their latest research progress or preliminary results in the RE community, especially considering the shorter review process of conferences and workshops. This is a common publication approach for relatively new and merging research areas. It also implies that RE of mobile apps is still an active topic and expects continuous attention from more researchers.  

In addition, Figure~\ref{fig_year} shows that the first study referring to RE of mobile apps was published in 2012. The reason could be that both Apple App Store and Google Play were launched in 2008, as mentioned in Section~\ref{study_search}, and several years are needed to make users and vendors provide various types of data to those apps. 

\subsection{Reflection on the Answer to RQ1}
First, regarding the sources of app data summarized in Table~\ref{tab_datasource1}, we found that app repositories were the main sources of app data for nearly 90\% of the included studies. We think, this could be explained with the following situation. First, in companies, the app developers should provide app descriptions, different releases, and app changelogs. for app users. Whereas, app users would like to send feedback in such an easy and convenient manner, after they used the app downloaded from this repository. Therefore, both app vendors and users contributed rich app data in app repositories. The other reason could be that as a family of mobile apps, it is much more convenient for researchers to collect various types of app data. Especially, some repositories provided specified APIs for this purpose. 

Besides app repositories, other sources, e.g. Twitter and Stack Overflow, served research purpose as well.  Although only eight out of the 43 included studies used such sources for exploration, it indicates that app repositories are the first but not the only choice for researchers to collect app data and form the datasets. Furthermore, it implies that app repositories have been overexploited, and new sources are sought after for research on RE of mobile apps.  

Next, app reviews were the main type of app data in the datasets for the RE purpose. We observed this in the distribution of app data types in Table~\ref{tab_datasource2}. This observation is consistent with findings in ~\cite{Ref2}. Moreover, other types of app data, such as app source code, twitters referring to users' feedback, app descriptions, were employed in a few primary studies. One reason for this could be that app reviews were reported as the typical explicit and textual user feedback~\cite{Ref2}. Compared to other types of app data, app reviews are not only easier to collect but more probable to imply user's requirements~\cite{Ref2}. Another reason could be that the text processing and analysis techniques are becoming mature to be applied in app reviews. 

Similarly, the exploration on the use of other types of app data indicates that the researchers and/or practitioners are interested in expanding the potential of existing sources well beyond app reviews. Therefore, we assume that the present research topic will remain active by supplementing more and more types of app data with app reviews, and then possibly attract more attention from researchers and/or practitioners.

\subsection{Reflection on the Answer to RQ2}

As observed in Section~\ref{RQ2}, this SMS reported RElic and RA as the two most popular RE activities, in which app data was employed. Specifically, all the 43 included studies mentioned RA, and around 93\% of the included studies reported RElic. Moreover, these two RE activities were usually reported in the same primary study. We think that the reasons could be the following. First, in most primary studies, the way of using the datasets referred to the analysis of app data, which can be further mapped to RA. Second, the app data needed to be pre-processed to extract and identify features manually or automatically before analysis, and RElic was involved in this process. 

Unlike RElic and RA, we found that RSp, RV and RMgt did not get enough researchers' attention. The reason could be that these three RE activities are relatively difficult to investigate by processing and analyzing the app datasets. Therefore, it remains to be seen to what extent the use of app datasets could support these three RE activities. 

\section{Limitations}
This SMS has some limitations. 
First, the primary studies were only selected from the automatic search in one digital library, i.e. IEEE, due to limited time spent on data extraction. Would the results differ if we explored more digital databases? There is a good chance that our findings would be similar, if other digital libraries, such as Scopus and Web of Science, are supplemented in the near future. 

Second,  both the selection and data extraction of primary studies depended on our RE knowledge. To avoid the possible bias and make the survey reproducible by other researchers or practitioners, we developed a study protocol to define our search strategy and specify the study selection procedure with IC and EC. Especially, to evaluate and improve the protocol, we conducted a pilot selection, and the first two authors were involved in the three round selection to assure a common understanding on study selection. Furthermore, the first two authors performed the data extraction on the result of pilot selection and generated a template to describe the data derived from primary studies and support our qualitative synthesis.

\section{Conclusions and Future Work}

This paper carried out a SMS on the state-of-the-art in the use of datasets for research in RE for mobile apps. Our most important conclusions are the following. First, there is a growth in the use of datasets for RE research since 2012 (see Figure~\ref{fig_year}). Second, despite the general interest in employing real world datasets in RE research, it seems the RE knowledge might be skewed due to the overuse of Google Play and App Store. We observed that there are attempts to supplement app reviews from repositories with other data sources. We consider this a positive development as we think it is important to expand the alternative sources and experiments with complimentary use of multiple sources. This is necessary, if the community wants more generalizable results. Third, we found that requirements elicitation and analysis have been the most researched RE activities by using datasets. While this is a positive finding, we call for more research in other RE activities, i.e. requirements management and validation.  

The next steps of this SMS include: (1) to add publications on this research topic from other digital libraries (e.g. Scopus and Web of Science); (2) to further explore the size of datasets used in RE of mobile apps as well as the manual/automatic methods employed to process those datasets for RE purpose.


\bibliographystyle{ACM-Reference-Format}
\bibliography{sample-base}

\appendix

\end{document}